\documentclass[10pt,twoside]{article}
\usepackage{amsfonts}
\usepackage{times}
\usepackage[utf8]{inputenc}
\usepackage[T1]{fontenc}
\usepackage{graphicx}
\usepackage{todonotes}
\usepackage{booktabs}
\usepackage{multirow}

\usepackage{siunitx}
\usepackage{hyperref}

\usepackage{coria-taln2026}
\usepackage[french]{babel} 
\usepackage{amsmath}
\usepackage{cleveref}

\usepackage{tikz}
\usetikzlibrary{arrows.meta, positioning, calc, shapes.geometric}
\usepackage{xcolor}

\definecolor{scblue}{RGB}{221,235,247}
\definecolor{scgreen}{RGB}{226,240,217}
\definecolor{scorange}{RGB}{252,228,214}
\definecolor{scgray}{RGB}{245,245,245}

\title{Sparse Coverage: Semantic Center Representations for Patent Prior-Art Retrieval}

\author{You Zuo\up{1,2} \quad Kim Gerdes\up{1,3} \quad  Éric de la Clergerie\up{2} \quad Beno\^{\i}t Sagot\up{2}\\
  {\small
    (1) Questel, Paris, France \\ 
    (2) Inria, Paris, France \\  
    (3) Laboratoire Interdisciplinaire des Sciences du Numérique, Université Paris-Saclay, Orsay, France \\ 
    \texttt{
      firstname.lastname@inria.fr, gerdes@lisn.fr \\ 
}}}

\begin{document}
\maketitle
\setlength{\parskip}{0.48\baselineskip}
\NoAutoSpacing

\resume{
Patent prior-art retrieval is a recall-oriented search task over long and highly structured technical documents. Dense retrieval improves semantic matching, but single-vector representations may compress multiple technical components, functions, and constraints into a single embedding. We propose Sparse Coverage, an unsupervised semantic retrieval framework that maps local span embeddings to a sparse vocabulary of embedding-space centers. The centers are selected with a coverage-oriented k-center objective, and spans activate nearby centers to produce sparse representations compatible with inverted-index retrieval. Experiments on CLEF-IP 2013 show that Sparse Coverage matches or exceeds the document-level recall of strong dense patent encoders in several configurations, while remaining competitive for passage-level retrieval. By combining local semantic evidence with sparse inverted-index search, Sparse Coverage provides an effective first-stage retrieval approach for patent search.
}

\abstract{Sparse Coverage : représentations par centres sémantiques pour la recherche de brevets}{
La recherche d’antériorité dans les brevets est une tâche orientée vers le rappel, portant sur des documents techniques longs et fortement structurés. La recherche dense améliore l’appariement sémantique, mais les représentations à vecteur unique peuvent comprimer plusieurs composants, fonctions et contraintes techniques dans un seul embedding. Nous proposons Sparse Coverage, un cadre non supervisé de recherche sémantique qui projette des embeddings locaux de segments textuels vers un vocabulaire creux de centres dans l’espace des embeddings. Ces centres sont sélectionnés au moyen d’un objectif k-center orienté vers la couverture, et les segments activent les centres voisins afin de produire des représentations creuses compatibles avec la recherche par index inversé. Les expériences menées sur CLEF-IP 2013 montrent que Sparse Coverage atteint ou dépasse, dans plusieurs configurations, le rappel au niveau document de puissants encodeurs denses de brevets, tout en restant compétitif pour la recherche au niveau passage. En combinant des indices sémantiques locaux avec une recherche creuse par index inversé, Sparse Coverage constitue une approche efficace pour la première étape de la recherche de brevets.
}

\motsClefs
  {document representation, patent, prior-art search, sparse retrieval}
  {représentation de documents, brevets, recherche d’antériorités, recherche creuse}





\section{Introduction}

Patent prior-art search is a core task in intellectual property workflows, supporting patent examination, invalidity analysis, and technology landscaping. Given a patent application, the objective is to retrieve earlier patents that disclose related technical ideas. The task is particularly challenging because patent documents are long, highly structured, and written in specialized technical language. In addition, prior-art search is strongly recall-oriented: missing a single relevant document may have significant legal and economic consequences~\cite{magdy2010pres,Bekkers2020TheIOA,deng2024impact}.

Recent neural retrieval models improve semantic matching beyond lexical overlap by encoding patents into dense embeddings~\cite{bekamiri2024patentsberta,vowinckel2023searchformer,ghosh2024paecter}. However, representing a long technical query or document with a single fixed-dimensional vector creates a capacity bottleneck: multiple components, functions, constraints, and alternatives must be compressed into one point in the embedding space. Recent theoretical and empirical work further suggests that fixed-dimensional single-vector retrieval has limited expressive capacity, since all relevance signals must be expressed through similarity to one query vector~\cite{weller2025theoretical}. This is especially relevant for patent prior-art search, where long claims combine multiple technical components, functions, and constraints. As patent collections grow, the candidate pool contains increasingly diverse near-relevant documents, making the required top-$k$ distinctions more complex.

Several neural retrievers move beyond this single-vector formulation in different ways. SPLADE~\cite{formal2021spladev2} keeps inverted-index retrieval by expanding over the pretrained encoder's tokenizer vocabulary, but its sparse dimensions remain fixed tokenizer terms. ColBERT~\cite{khattab2020colbert,santhanam2022colbertv2} preserves fine-grained matching by retaining token-level dense vectors and using MaxSim interactions, but requires specialized dense-token indexing and scoring. We instead seek a lighter first-stage semantic retriever that converts local dense evidence into sparse embedding-space dimensions, with capacity controlled by vocabulary size $V$.

In this work, we introduce \textbf{Sparse Coverage}, an unsupervised retrieval framework that relaxes the single-vector bottleneck while preserving inverted-index search. Instead of compressing all technical components of a patent claim or passage into one dense point, Sparse Coverage represents text through local semantic regions activated by contextual span embeddings. Patent texts are decomposed into encoder tokens, noun phrases, or hybrid span units and embedded with a pretrained encoder. We construct a semantic vocabulary by selecting $V$ observed span embeddings as \emph{centers} that cover the span-embedding space; each center is anchored by an actual text span and acts as a dictionary-like semantic unit. At encoding time, each span activates only the few centers whose learned radii cover it, yielding sparse vectors over embedding-space regions. The vocabulary size $V$ explicitly controls representation granularity and capacity: larger vocabularies define finer regions and preserve more local relevance signals. Retrieval is performed by inverted-index traversal over shared activated centers, with matches traceable to the query and document spans that activated them.

We evaluate Sparse Coverage on the CLEF-IP 2013 prior-art benchmark for both document- and passage-level retrieval~\cite{Piroi2013PassageRSA}. We compare against lexical retrieval, dense patent encoders, neural sparse retrieval, and late-interaction baselines, and analyze how retrieval behavior changes with the vocabulary size $V$. Experiments show that Sparse Coverage provides a strong document-level recall and retrieval-time scoring trade-off by replacing exhaustive dense query--candidate scoring with selective traversal over activated center postings.
Code and resources are available for reproducibility.\footnote{\url{https://github.com/ZoeYou/patent-sparse-coverage}}

\section{Related Work}
\label{sec:related_work}


\paragraph{Patent retrieval.}
Patent prior-art search has received sustained attention due to its practical importance and the availability of benchmarks derived from patent examination reports. The CLEF-IP campaigns released large-scale EPO collections and established benchmarks for document-level and claim-to-passage retrieval~\cite{Roda2009CLEFIP2RA,piroi2012clef,Piroi2013PassageRSA}.

Early systems mainly relied on sparse retrieval with query engineering and structure-aware heuristics, including field-based lexical matching, keyphrase extraction, query expansion, and filtering with patent metadata such as IPC codes~\cite{verma2011exploring,bouadjenek2015study}. Prior work also showed that noun phrases are useful retrieval units for patent passage retrieval~\cite{andersson2013exploring}. This aligns with patent language, where core inventive content is often expressed through noun phrases denoting technical entities, components, materials, and functional units.

Recent work has introduced patent-oriented neural retrieval models and encoders, including PatentSBERTa~\cite{bekamiri2024patentsberta}, citation-supervised models such as PaECTER~\cite{ghosh2024paecter}, SearchFormer~\cite{vowinckel2023searchformer}, and QaECTER~\cite{djemmal2026citation}, multi-task models such as PatEmbed~\cite{ayaou2025patenteb}, and self-supervised patent encoders based on intra-document views~\cite{zuo2025patent}. These models provide strong semantic representations for patents and serve as important dense baselines in our experiments.

\paragraph{Neural retrieval beyond single-vector embeddings.}
Single-vector dense retrieval is only one way to use neural representations for search. Neural sparse models such as SPLADE~\cite{formal2021spladev2} and uniCOIL~\cite{gao2021coil} retain inverted-index retrieval while learning contextual term weighting or expansion over the encoder vocabulary; SparseEmbed~\cite{kong2023sparseembed} further combines sparse lexical activations with contextual embeddings. These methods improve lexical sparse retrieval, but their dimensions remain tied to a fixed term vocabulary.

A second alternative is late interaction. ColBERT~\cite{khattab2020colbert,santhanam2022colbertv2} stores token-level dense representations and scores query--document pairs through MaxSim interactions. This preserves local matching evidence that single-vector encoders may lose, while requiring specialized indexing and scoring for dense token vectors.

Closest to our goal, \cite{dobrynin2025efficient} build an embedding-based inverted index by clustering contextualized token embeddings into an extended vocabulary. Sparse Coverage also builds an inverted index over embedding-derived units, but instead of clustering meanings within lexical tokens, it chooses centers that geometrically cover the span-embedding space. These centers act as dictionary-like semantic units optimized for retrieval-oriented coverage rather than reconstruction, making representation capacity adjustable via the vocabulary size $V$ while avoiding a fixed lexical term space.




\begin{figure}[t]
\centering
\begin{tikzpicture}[
    font=\footnotesize,
    >=Latex,
    box/.style={
        draw=black!70,
        rectangle,
        minimum height=0.6cm,
        minimum width=1.7cm,
        align=center,
        fill=white,
        inner sep=2pt
    },
    proc/.style={
        box,
        draw=violet!65,
        fill=violet!10,
        minimum width=2.0cm
    },
    voc/.style={
        box,
        draw=blue!70!black,
        fill=cyan!15,
        minimum width=2.2cm,
        minimum height=0.85cm
    },
    dia/.style={
        draw=violet!65,
        diamond,
        aspect=1.1,
        minimum width=1.2cm,
        minimum height=1.0cm,
        align=center,
        fill=violet!10,
        inner sep=1pt
    },
    arr/.style={->, thick, black!70},
    refarr/.style={->, thick, dotted, black!65}
]

\filldraw[
    fill=yellow!7,
    draw=olive!45,
    thick
] (-1.9, -1.65) rectangle (7.0, -0.35);

\node[font=\bfseries\footnotesize] at (-1.1, -0.55) {Offline};

\filldraw[
    fill=yellow!7,
    draw=olive!45,
    thick
] (-0.6, 0.35) rectangle (12.0, 1.65);

\node[font=\bfseries\footnotesize] at (0.2, 1.5) {Online};

\node[box]  (corpus)  at (-0.8, -1.0) {Corpus};
\node[proc] (spanenc) at (1.4, -1.0) {Span\\Embeddings};
\node[voc] (semvoc) at (4.8, -1.0) {Semantic Vocabulary\\(Centers, Radii)};

\draw[arr] (corpus) -- (spanenc);
\draw[arr] (spanenc) -- node[midway, above, font=\scriptsize] {$k$-center} (semvoc);

\node[box, minimum height=0.75cm, minimum width=1.9cm]
    (qdoc) at (0.6, 1.0) {Query / New\\Document};
\node[proc, minimum width=2.0cm] (onspan) at (2.9, 1.0) {Span\\Embeddings};
\node[proc, minimum width=2.4cm] (activate) at (5.5, 1.0) {Radius + Top-$K$\\Activation};
\node[voc, minimum width=2.0cm, minimum height=0.65cm] (sparse) at (8.1, 1.0) {Sparse Center\\Vector};
\node[proc, minimum width=2.2cm, minimum height=0.75cm]
    (retrieval) at (10.6, 1.0) {Inverted Index\\Retrieval};

\draw[arr] (qdoc) -- (onspan);
\draw[arr] (onspan) -- (activate);
\draw[arr] (activate) -- (sparse);
\draw[arr] (sparse) -- (retrieval);

\draw[refarr]
    (semvoc.north) to[out=90, in=270] 
    node[midway, right, font=\scriptsize, align=left] {shared\\vocabulary}
    (activate.south);

\end{tikzpicture}
\caption{Overview of Sparse Coverage. A semantic vocabulary of centers and radii is built offline from span embeddings, then reused to map queries and documents into sparse center activations for inverted-index retrieval.}
\label{fig:demo}
\end{figure}
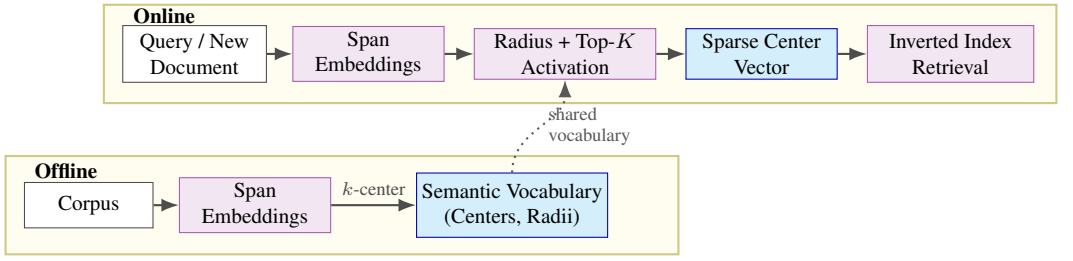

\section{Methodology}
\subsection{Semantic Center Construction}
Figure~\ref{fig:demo} illustrates the offline--online pipeline of Sparse Coverage. The semantic vocabulary is constructed offline from a large patent corpus and then reused at retrieval time to encode queries and corpus documents as sparse center activations.

Let $\mathcal{D}_{\mathrm{vocab}}=\{d_1,\dots,d_N\}$ denote the corpus used for vocabulary construction. We assume a pretrained encoder $f_\theta$ that maps text to contextual token embeddings. Each document section is encoded as a sequence, from which textual spans are extracted. We instantiate textual spans in three ways: encoder tokens, noun phrases, and a hybrid unit consisting of noun phrases plus encoder tokens not covered by any noun phrase. Each span is represented by mean pooling over its contextual token embeddings, followed by $\ell_2$ normalization. This yields a set of normalized span embeddings $\mathcal{X}=\{x_1,\dots,x_T\}$, where $x_i\in\mathbb{R}^d$ and similarity is measured by cosine similarity $\langle x,y\rangle$. Normalization ensures that center selection and activation depend on angular similarity rather than embedding norms.

\paragraph{Center selection.}
We construct a semantic vocabulary by selecting $V$ span embeddings from $\mathcal{X}$ as semantic centers, denoted by $C=\{c_1,\dots,c_V\}\subset\mathcal{X}$. The vocabulary size $V$ controls the granularity of the semantic covering. To encourage broad coverage of the span embedding space, we use the following $k$-center objective:
\begin{equation}
\min_{|C|=V} \max_{x \in \mathcal{X}} \min_{c \in C} (1-\langle x,c\rangle).
\label{eq:kcenter}
\end{equation}
This worst-case objective matches the high-recall goal of prior-art search: rare or peripheral semantic regions should not be ignored.
In practice, we use farthest-first traversal~\cite{gonzalez1985clustering}, a standard greedy procedure for the $k$-center problem.

\paragraph{Center-specific radii.}
After center selection, each span is assigned to its nearest center:
\begin{equation}
a(x)=\arg\min_{c\in C}(1-\langle x,c\rangle).
\label{eq:assignment}
\end{equation}
For each center $c$, we define a center-specific coverage radius from the distances of spans in its Voronoi cell:
\begin{equation}
r_c = Q_\tau \left(\{1-\langle x,c\rangle : a(x)=c\}\right),
\label{eq:radius}
\end{equation}
where $Q_\tau$ denotes the $\tau$-th percentile. 
Using a high percentile rather than the maximum reduces sensitivity to outliers while allowing dense and dispersed semantic regions to receive different activation thresholds.

\subsection{Sparse Retrieval via Center Activations}
Given centers $C$ and radii $\{r_c\}$, Sparse Coverage maps each span embedding to a small set of activated centers. These activations define sparse query and corpus-unit representations over the semantic vocabulary, where a corpus unit may be a document or a passage depending on the retrieval setting.

\paragraph{Center activation.}
For a span embedding $x$, candidate centers are those whose center-specific radius covers the span:
\begin{equation}
\mathcal{A}(x)=
\operatorname{TopK}_K\{c\in C \mid 1-\langle x,c\rangle \le r_c\},
\label{eq:activation}
\end{equation}
where $\operatorname{TopK}_K$ retains the $K$ centers with largest $\langle x,c\rangle$ if more than $K$ candidates satisfy the radius condition. The radius condition defines an adaptive semantic neighborhood, whereas the top-$K$ cap is a sparsification mechanism. It prevents spans located in dense regions of the embedding space from activating too many overlapping centers.

\paragraph{Center-level aggregation.}
For a corpus unit $d$ with spans $S_d=\{x_i\}$, the weight of center $c$ is
\begin{equation}
w_{d,c}=\max_{x_i\in S_d}\mathbf{1}[c\in\mathcal{A}(x_i)]\langle x_i,c\rangle.
\label{eq:wdc}
\end{equation}
That is, a center receives the maximum similarity over all spans in $d$ that activate it.
This max aggregation preserves strong semantic matches while reducing the influence of repeated weak activations. To reduce the bias toward longer corpus units, which contain more spans and therefore have more opportunities to activate centers, we apply corpus-side span-count normalization, $\tilde{w}_{d,c}=w_{d,c}/|S_d|^\gamma$. Queries are encoded and aggregated in the same center vocabulary, but this length normalization is applied only to corpus units.

\paragraph{Sparse interaction scoring.}
Retrieval is performed with an inverted index over centers. After center aggregation, we compute the document frequency of each center,
$
\mathrm{df}_c = |\{d \in \mathcal{D}: w_{d,c}>0\}|,
$
where $\mathcal{D}$ denotes the indexed corpus units. A small fraction of the most frequent centers are treated as stop centers: we suppress the top $\rho$ fraction of centers with the largest $\mathrm{df}_c$ from retrieval scoring. These centers correspond to very generic semantic regions, produce long posting lists, and contribute little discriminative evidence, similarly to stopwords in lexical retrieval.

For each remaining center $c$, we define
$\mathrm{idf}_c=\log\frac{N+1}{\mathrm{df}_c+1}+1.$
Although the $k$-center objective encourages geometric coverage, it operates over spans rather than documents. As a result, geometric coverage does not imply balanced document-frequency statistics: multiple spans from the same document may influence center construction, and some generic semantic regions may still be activated across many corpus units. We therefore compute document frequency after center activation, suppress the most frequent centers, and use IDF weighting to calibrate the contribution of the remaining centers at retrieval time. 
This document-frequency imbalance is further analyzed in Appendix~\ref{app:center-analysis} through activation-sparsity and posting-list-skew measurements.

Given a query $q$ and a corpus unit $d$, the final score is
\begin{equation}
\mathrm{score}(q,d)=
\sum_{c\in(\mathrm{supp}(q)\cap\mathrm{supp}(d))\setminus C_{\mathrm{stop}}}
w_{q,c}\,\tilde{w}_{d,c}\,\mathrm{idf}_c^\alpha,
\label{eq:score}
\end{equation}
where $C_{\mathrm{stop}}$ is the set of suppressed stop centers. Only non-stop centers activated by both the query and the corpus unit contribute to Eq.~\eqref{eq:score}. This makes retrieval compatible with standard inverted-index traversal while allowing semantic matching through shared embedding-space centers.

\section{Experimental Settings}
\subsection{Data for Center Construction}
We build the semantic vocabulary from English patent documents in the European Patent Office (EPO) bulk data.\footnote{\url{https://www.epo.org/en/searching-for-patents/data/bulk-data-sets/data}} The full English EPO collection from 2000--2010 contains roughly 2M documents. From this collection, we sample 300{,}000 documents proportionally by publication year to preserve the original temporal distribution.

Each patent is divided into \textit{title+abstract}, \textit{claims}, and \textit{description} sections. We build vocabularies from three span units: \textit{encoder tokens}, \textit{noun phrases}, and a \textit{hybrid noun-phrase--token} unit containing noun phrases plus remaining encoder tokens. Noun phrases are extracted with SciSpaCy (\texttt{en\_core\_sci\_lg})~\cite{neumann2019scispacy}; encoder tokens are obtained from the corresponding transformer tokenizer. Because the total number of extracted spans is very large, we retain up to 5M spans per unit, sampled proportionally across the three sections. Center selection is performed on this sampled span set, and center-specific radii are estimated from the resulting Voronoi assignments.

\subsection{Evaluation Setup}
\paragraph{Dataset.}

\begin{figure}[ht]
    \centering
    \includegraphics[width=0.85\columnwidth]{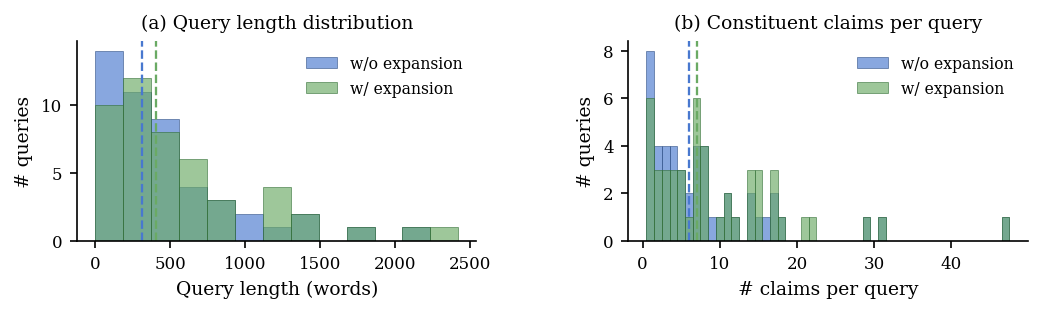}
    \caption{Distribution of CLEF-IP 2013 query lengths, showing substantial variation in both words per query (left) and claims per query (right).}    
    \label{fig:query-length-stats}
\end{figure}

We evaluate document- and passage-level retrieval on the CLEF-IP 2013 claims-to-passages benchmark~\cite{Piroi2013PassageRSA}. Queries are patent claims, and relevance is annotated at the passage level from EPO search reports. We restrict evaluation to English patents, yielding 48 usable queries.\footnote{The benchmark defines 50 English topics (PSG-1--PSG-50), but two topics (PSG-3 and PSG-4) point to \texttt{EP-1487746-A2}, whose XML contains no \texttt{<claims>} element; the corresponding WO family documents are also absent from the distributed \texttt{tfiles}. We therefore exclude these two topics.}
For dependent-claim queries, we prepend ancestor independent claims in claim-number order so that the query is self-contained and consistent with prior-art search practice.

After preprocessing, each query has about 3.4 relevant documents and 67.5 relevant passages on average. We first construct a 25{,}000-document English candidate pool by retaining all 90 relevant documents and sampling negative documents, with 75\% IPC-based hard negatives (\emph{subgroup-hard}, \emph{maingroup-hard}, and \emph{subclass-hard}) and 25\% uniformly sampled negatives. After XML parsing and passage extraction, the realized evaluation corpus contains 17{,}323 usable documents and 1.4M passages. All extractable passages from these documents are kept. Query length and number of constituent claims vary substantially across topics; Figure~\ref{fig:query-length-stats} summarizes these distributions. Additional statistics and qualitative examples are provided in Appendix~\ref{app:clefip2013} and~\ref{app:clefip-examples}.

\paragraph{Protocol and metrics.}
We use a two-stage evaluation pipeline. First, passages are ranked and deduplicated by document, using each document's highest-ranked passage as its first occurrence. The top-\(K\) unique documents (\(K=100\)) are used for cutoff-based document metrics and define the candidate set for passage reranking. Second, all passages from these top-\(K\) documents are pooled and re-ranked by their original retrieval scores.

For document-level retrieval, we report Recall@100, mAP, and PRES@100~\cite{magdy2010pres}. Recall@100 and PRES@100 are computed on the top-100 document ranking, while mAP is computed on the full deduplicated document ranking. PRES@100 is defined as 
\(\mathrm{PRES@100}=1-\left(\frac{1}{R}\sum_{i=1}^{R} r_i-\frac{R+1}{2}\right)/N_{\max}\), 
where \(R\) is the number of relevant documents, \(N_{\max}=100\), and \(r_i\) is the actual or penalty rank of the \(i\)-th relevant document.
For passage-level retrieval, we report the official CLEF-IP mAP(D)~\cite{piroi2012clef}. For each query \(q\), \(AP(D_i;q)\) is computed on the subsequence of the passage ranking belonging to the relevant document \(D_i\), and mAP(D) averages these AP values over relevant documents and queries:
\vspace{-.2em}
\begin{equation}
\mathrm{mAP(D)}=\frac{1}{|Q|}\sum_{q\in Q}\frac{1}{n(q)}
\sum_{i=1}^{n(q)} AP(D_i;q).
\end{equation}
\vspace{-2.5em}

\subsection{Baseline Models}
We compare Sparse Coverage with lexical, dense, neural sparse, and late-interaction baselines. Since patent claims often exceed the 512-token limit of transformer encoders, we use sentence-aligned query chunking when needed. Passage texts are encoded independently.

\textbf{Lexical Retrieval.}
We use BM25 with stemming and stopword removal. Each claim is used in full.

\textbf{Dense Retrieval.}
We evaluate four fixed-vector dense baselines: SPECTER2~\cite{singh2022scirepeval}, BERT-for-Patents~\cite{SrebrovicYonamine2020}, PaECTER~\cite{ghosh2024paecter}, and PatentMap-V0-SecPair-Claim~\cite{zuo2025patent}.\footnote{We use the following HuggingFace checkpoints: \texttt{allenai/specter2}, \texttt{anferico/bert-for-patents}, \texttt{mpi-inno-comp/paecter}, and \texttt{ZoeYou/PatentMap-V0-SecPair-Claim}. For SPECTER2, we use its retrieval adapter on top of the SPECTER2 base model.} Corpus passages are paragraph-level units and are encoded independently as fixed-dimensional embeddings. Since patent claims often exceed the input length of transformer encoders, long claims are split into sentence-aligned chunks of up to 512 tokens using each model's tokenizer. Each query chunk is encoded independently, and its similarity to a passage is computed by cosine similarity. The final query--passage score is the maximum cosine similarity over all query chunks. Passages are then ranked according to this score. 

\textbf{Neural Sparse and Late Interaction.}
We include SPLADE-v2~\cite{formal2021spladev2} and ColBERT-v2~\cite{santhanam2022colbertv2} as representative off-the-shelf neural sparse and late-interaction baselines; they are not further adapted to the patent domain.  
For SPLADE, we apply the same query chunking as for dense baselines and merge chunk-level sparse vectors by elementwise maximum. 
For ColBERT-v2, we use the \texttt{colbert-ir/colbertv2.0} checkpoint with the PLAID engine. 
Corpus passages are indexed as independent passages with document length capped at 512 tokens. 
Long claim queries are split into chunks at search time; chunks are searched independently and passage scores are merged by maximum score across chunks.

\subsection{Implementation Details}
Unless otherwise stated, Sparse Coverage uses soft assignment with up to $K=5$ centers per span, suppresses the top $1\%$ most frequent centers as stop centers, applies max aggregation for center weights, uses IDF weighting with $\alpha=2.0$, and applies corpus-side span-count normalization $\tilde{w}_{d,c}=w_{d,c}/|S_d|^\gamma$ with $\gamma=0.5$. At inference time, long claims exceeding the encoder limit are split into sentence-aligned 512-token chunks. Each chunk is encoded independently, but unlike dense chunk-level max scoring, Sparse Coverage merges all extracted query spans into a single sparse center representation before retrieval.

We evaluate vocabulary sizes $V \in \{10k, 20k, \dots, 50k\}$ and construct vocabularies for three span granularities: encoder tokens (enc. tok.), noun phrases (NP), and a hybrid unit consisting of noun phrases plus remaining encoder tokens (NP+tok.). To test robustness across embedding spaces, we use BERT-for-Patents~\cite{SrebrovicYonamine2020}, PaECTER~\cite{ghosh2024paecter}, and PatentMap-V0~\cite{zuo2025patent}. Unless otherwise noted, all reported results use this default setting. Appendix~\ref{sec:ablation_study} examines assignment strategy, IDF exponent, and normalization strength.

\section{Results}
\subsection{Retrieval Effectiveness}
\label{sec:main_result}

\begin{table*}[ht]
\centering
\caption{Main retrieval results on CLEF-IP 2013. For Sparse Coverage, we report selected strong configurations: the best document-level Recall@100 setting for each encoder, plus the best passage-level MAP(D) setting. 
We use this selected view only to compare attainable operating points; systematic trends over all configurations are analyzed in Section~\ref{sec:granularity} and Appendix~\ref{app:full-results}.}

\label{tab:main-results}
\footnotesize
\setlength{\tabcolsep}{5pt}
\begin{tabular}{llcccccc}
\toprule
\multirow{2}{*}{\textbf{Model}} & \multirow{2}{*}{\textbf{Unit}} & \multirow{2}{*}{$V/d$} 
& \multicolumn{3}{c}{\textbf{Document-level}} 
& \multicolumn{2}{c}{\textbf{Passage-level}} \\
\cmidrule(lr){4-6} \cmidrule(lr){7-8}
& & & \textbf{R@100} & \textbf{mAP} & \textbf{PRES@100} & \textbf{MAP(D)} & \textbf{R@1000} \\
\midrule
\multicolumn{8}{l}{\textit{Lexical Retrieval}} \\
BM25 & -- & -- & 85.90 & 46.62 & 72.14 & 25.91 & 38.20 \\
\midrule

\multicolumn{8}{l}{\textit{Neural Sparse \& Late Interaction}} \\
SPLADE-v2 & -- & 30,522 & 89.51 & 47.57 & 80.46 & 24.34 & 42.15 \\
ColBERTv2 & -- & 128 & 71.28 & 40.19 & 64.82 & 19.21 & 16.16 \\
\midrule

\multicolumn{8}{l}{\textit{Dense Retrieval}} \\
SPECTER2 & -- & 768 & 82.01 & 41.18 & 71.94 & 21.52 & 34.28 \\
BERT-for-Patents & -- & 1,024 & 66.08 & 33.02 & 57.66 & 19.45 & 28.59 \\
PaECTER & -- & 1,024 & 97.22 & \textbf{58.61} & \textbf{90.33} & 26.63 & \underline{54.05} \\
PatentMap-V0 & -- & 1,024 & 96.53 & \underline{57.01} & \underline{87.05} & 26.68 & \textbf{56.72} \\
\midrule

\multicolumn{8}{l}{\textit{Sparse Coverage (Ours): selected configurations}} \\
BERT-for-Patents & enc. tok. & 30k & 95.21 & 46.70 & 83.46 & \underline{29.54} & 44.17 \\
BERT-for-Patents & enc. tok. & 40k & 93.47 & 46.05 & 82.84 & \textbf{30.03} & 45.32 \\
PaECTER & NP+tok. & 50k & \textbf{99.31} & 45.44 & 85.21 & 25.35 & 45.37 \\
PatentMap-V0 & NP & 20k & \underline{97.71} & 30.90 & 81.29 & 25.40 & 43.52 \\
\bottomrule
\end{tabular}
\end{table*}

Since prior-art search is recall-oriented, Table~\ref{tab:main-results} highlights Sparse Coverage configurations that define its strongest operating points: the highest document-level Recall@100 setting for each encoder and the overall highest passage-level MAP(D) setting. Full results across all vocabulary sizes and span units are reported in Appendix~\ref{app:full-results}.

Sparse Coverage achieves the highest document-level Recall@100. 
With PaECTER embeddings, NP+tok. units, and $V=50\mathrm{k}$, it reaches 99.31 Recall@100, exceeding the strongest dense baseline and the off-the-shelf neural sparse and late-interaction baselines in our setup. PatentMap-V0 with noun-phrase units also reaches 97.71 Recall@100, slightly above its dense counterpart. These results support the use of Sparse Coverage as a high-recall first-stage retriever: local span evidence mapped to a sparse semantic vocabulary can recover relevant prior-art documents that are not always captured by a single dense vector.

In its best passage-oriented configuration, Sparse Coverage also obtains the highest passage-level MAP(D). BERT-for-Patents with encoder-token units reaches 30.03 MAP(D) at $V=40\mathrm{k}$, while the nearby $V=30\mathrm{k}$ setting already reaches 29.54. This is notable because the dense BERT-for-Patents baseline performs poorly on the same benchmark. Converting it into Sparse Coverage yields more than 27 absolute points in document Recall@100 and more than 10 absolute points in MAP(D), corresponding to relative improvements above 40\% and 50\%, respectively. This suggests that BERT-for-Patents contains useful local semantic signals, but these signals are largely lost when passages are represented by a single pooled vector.

The improvements are not uniform across all ranking metrics. Dense PaECTER and PatentMap-V0 remain stronger on document-level mAP and PRES@100, and PatentMap-V0 obtains the highest passage Recall@1000. This indicates that dense patent encoders still provide better score calibration for ranking some relevant items early. Sparse Coverage is therefore best viewed as a sparse semantic retriever optimized for high-recall candidate generation and, in some settings, improved document-conditioned passage ranking, rather than as a universal replacement for dense ranking.

\subsection{Span Granularity and Vocabulary Size}
\label{sec:granularity}

\begin{figure}[ht]
\centering
\includegraphics[width=\linewidth]{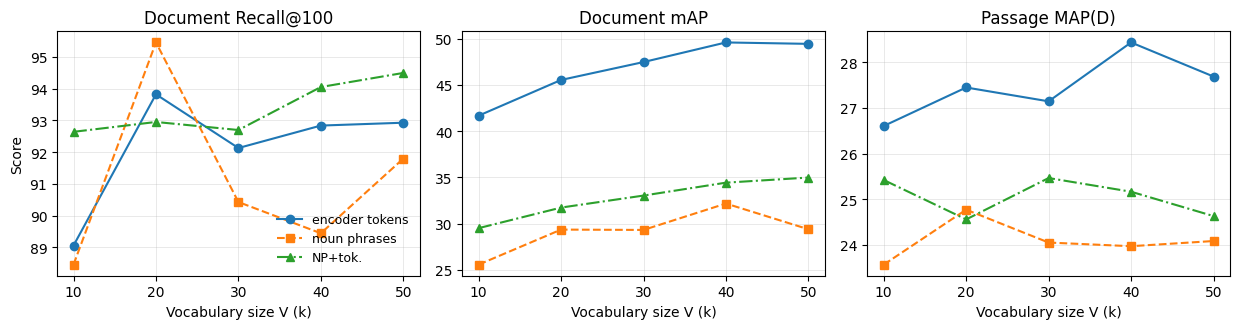}
\caption{
Effect of vocabulary size and span granularity on Sparse Coverage.
Each curve averages over the three encoders. Increasing the vocabulary size generally improves document mAP, while its effect on document Recall@100 and passage MAP(D) is less monotonic. Span granularity produces clearer differences: encoder-token units are strongest for ranking-oriented metrics, whereas the hybrid NP+tok. unit provides strong recall-oriented coverage.
}
\label{fig:vocab-size}
\end{figure}

Figure~\ref{fig:vocab-size} summarizes the effect of vocabulary size and span granularity by averaging Sparse Coverage results over the three encoders. This analysis complements Table~\ref{tab:main-results}: while Table~\ref{tab:main-results} reports selected strong configurations, Figure~\ref{fig:vocab-size} shows the aggregate behavior across all tested values of $V$ and all span units. The full non-averaged results for all encoder--unit--vocabulary combinations are reported in Appendix~\ref{app:full-results}.

The vocabulary size $V$ controls the resolution of the semantic covering. On average, document Recall@100 peaks at an intermediate vocabulary size for encoder-token and noun-phrase units, whereas the hybrid NP+tok. unit is more stable and benefits from larger vocabularies, showing that the best resolution depends on the span unit. Increasing $V$ creates finer centers, improving match specificity and often helping early document ranking, as reflected by the upward trend in document mAP. However, finer partitions also reduce overlap between related spans: semantically close query and document spans may activate neighboring but distinct centers, weakening sparse matches. This can affect all metrics, but is especially visible in Recall@100, which depends on preserving enough shared activations to recover relevant documents within the top 100.

Span granularity has a clearer and more consistent effect. Encoder-token units achieve the highest average document mAP and passage MAP(D), as well as the best passage-level MAP(D) overall, showing that fine-grained local evidence is important for early ranking and passage-level retrieval. Noun-phrase units are weaker on average for ranking-oriented metrics, but they remain useful for document-level recall; for example, PatentMap-V0 with noun-phrase units reaches 97.71 Recall@100 at $V=20k$. The hybrid NP+tok. unit combines concept-level noun phrases with remaining token-level evidence and reaches the best document-level Recall@100 overall, but it does not consistently improve ranking metrics over encoder-token units. Overall, span granularity should be chosen according to the retrieval objective: token-level units are preferable for ranking-oriented metrics, whereas hybrid or phrase-aware units are effective for high-recall first-stage retrieval. Additional geometric analysis of the induced vocabularies is provided in Appendix~\ref{app:center-analysis}.

\subsection{Efficiency Analysis}
\label{sec:efficiency_analysis}

\begin{figure}[ht]
\centering
\includegraphics[width=0.7\linewidth]{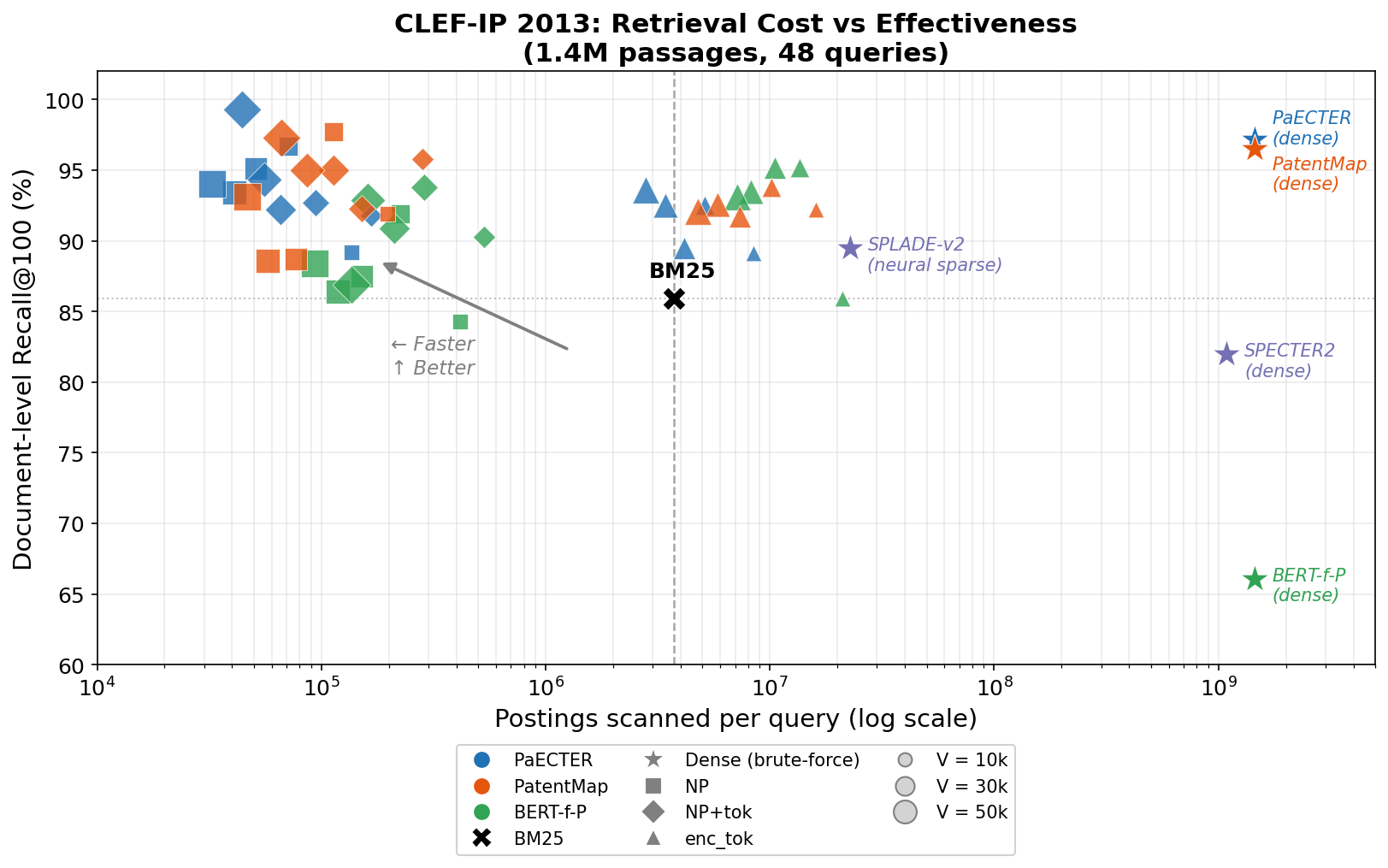}
\caption{
Retrieval cost vs.\ document-level Recall@100 on CLEF-IP 2013.
The x-axis reports the number of postings scanned per query for sparse methods
(log scale). For dense baselines, the marker indicates the number of query--candidate
similarity computations under exhaustive scoring in our evaluation setting.
}
\label{fig:efficiency_scatter}
\end{figure}

Figure~\ref{fig:efficiency_scatter} compares document-level Recall@100 with retrieval-time interaction cost. For Sparse Coverage and BM25, cost is measured as the number of postings scanned per query. For dense baselines, we report the number of query--candidate similarity computations under exhaustive scoring in our evaluation setting.
ColBERT-v2 is excluded from this cost plot because its PLAID late-interaction cost is not directly comparable to sparse posting traversal or fixed-vector exhaustive scoring. 
The dense-baseline cost should be interpreted as a reference cost, not as the best possible deployment cost of dense retrieval, since approximate nearest-neighbor indexes and vector compression can reduce dense search costs. The comparison is intended to highlight the retrieval mechanism of Sparse Coverage: semantic matching is performed through activated center posting lists, making the method compatible with standard inverted-index traversal. This analysis therefore measures retrieval-time scoring interactions after query encoding, rather than end-to-end latency; measuring full indexing and query-processing costs is left for future work.

Sparse Coverage occupies a favorable cost--recall region. Many configurations achieve higher Recall@100 than BM25 while scanning substantially fewer postings, and the best configurations reach recall levels comparable to, or higher than, strong dense patent retrievers without exhaustive dense scoring. This suggests that center activations provide a selective sparse access pattern: only documents sharing query-active semantic centers are scored, while the center activations still originate from contextual span embeddings produced by pretrained encoders.

The cost of Sparse Coverage is governed by two opposing effects. Increasing the vocabulary size $V$ creates finer semantic regions and usually shortens individual posting lists. However, queries contain many spans, and each span may activate several centers, increasing the number of lists accessed. The scanned postings therefore depend on the balance between posting-list length and query sparsity, explaining why cost is not monotonic in $V$ and why larger vocabularies can sometimes improve recall without increasing scanned postings.

\section{Conclusion}
We presented Sparse Coverage, an unsupervised framework that maps local patent span embeddings to sparse semantic center activations for inverted-index retrieval. Experiments on CLEF-IP 2013 show that Sparse Coverage is an effective first-stage retriever, matching or exceeding the document-level recall of strong dense patent encoders in several configurations while remaining competitive for passage-level retrieval. By representing patents through shared center activations, the method combines local semantic evidence with sparse index-based candidate selection. Our analysis shows that both span granularity and vocabulary size shape retrieval behavior: token-level units improve fine-grained ranking, phrase-aware units support high-recall coverage, and larger vocabularies can improve ranking-oriented metrics by increasing semantic resolution.

\section*{Acknowledgments}
We thank the anonymous reviewers for their insightful comments, the CLEPS infrastructure at Inria Paris for computational resources, and Younes Djemmal and Kirian Guiller for their helpful discussions and feedback.
This work was partly funded by the last author’s chair in the PRAIRIE institute funded by the French national agency ANR as part of the “Investissements d’avenir” programme under the reference ANR-19-P3IA-0001.
\bibliographystyle{coria-taln2026}
\bibliography{biblio}

\newpage
\appendix

\section{Statistics of Evaluation Data}
\label{app:clefip2013}

\begin{table}[ht]
\centering
\caption{CLEF-IP 2013 English sampled corpus statistics under the default evaluation setting. The sampler initially aimed to draw a 25{,}000-document pool with all relevant documents, 75\% IPC-based hard negatives, and 25\% random negatives. After XML parsing and passage extraction, the realized usable corpus contains 17{,}323 documents. Query lengths are measured after ancestor expansion of dependent claims.}
\label{tab:dataset-stats}
\small
\begin{tabular}{l r}
\toprule
\textbf{Statistic} & \textbf{Value} \\
\midrule
\multicolumn{2}{l}{\textit{Corpus}} \\
\quad Total documents                & 17{,}323 \\
\quad \quad Relevant documents       & 90 \\
\quad \quad Non-relevant documents   & 17{,}233 \\
\quad Total passages                 & 1{,}414{,}683 \\
\quad \quad From relevant docs       & 9{,}117 \\
\quad \quad From non-relevant docs   & 1{,}405{,}566 \\
\quad Passages per document (mean / median) & 81.7 / 59 \\
\midrule
\multicolumn{2}{l}{\textit{Queries}} \\
\quad Number of queries              & 48 \\
\quad Query length in words, expanded (mean / median) & 585 / 407 \\
\midrule
\multicolumn{2}{l}{\textit{Relevance}} \\
\quad Relevant passages per query (mean / median) & 67.5 / 44.0 \\
\quad Relevant documents per query (mean / median) & 3.4 / 3.0 \\
\quad Total relevant (query, passage) judgments & 3{,}239 \\
\quad Unique relevant passages (present in sample) & 1{,}752 / 1{,}834 \\
\midrule
\multicolumn{2}{l}{\textit{Passage length}} \\
\quad Mean / median (words)          & 84.1 / 58.0 \\
\bottomrule
\end{tabular}
\end{table}

Table~\ref{tab:dataset-stats} summarizes the realized CLEF-IP 2013 English evaluation corpus used in our default setting. The sampling script initially aimed to draw a 25,000-document pool, consisting of all relevant documents together with IPC-based hard negatives and random negatives. After XML parsing and passage extraction, the final usable corpus contains 17,323 unique documents and about 1.4M passages. The reduction is not due to a lack of IPC-matched negative candidates: before XML filtering, the sampler had access to large negative pools in each stratum. Instead, the reduction occurs after sampling, because some selected patents cannot be parsed or contain no extractable title, abstract, claims, or description passages.

The realized benchmark remains highly imbalanced, as expected in prior-art search. Only 90 of the 17,323 documents are relevant, and their 9,117 passages are mixed with more than 1.4M passages from non-relevant documents. Each query has on average only 3.4 relevant documents, so the document-level task evaluates whether the system can recover a small set of critical prior-art documents from a large candidate pool. This motivates recall-oriented metrics such as Recall@100 and PRES@100. The table also reports passage-level relevance statistics for completeness.

The queries are also long and structurally complex. After ancestor expansion of dependent claims, the mean query length is 585 words, with a median of 407 words. This confirms that the benchmark differs from standard short-query retrieval: each query may contain several technical constraints, components, and functional relations. These characteristics motivate our use of span-level representations and our evaluation of document-level recall.

\section{Examples of CLEF-IP 2013 Data}
\label{app:clefip-examples}

To provide a concrete understanding of the evaluation task described in Appendix~\ref{app:clefip2013}, we present representative examples of queries and their corresponding ground-truth relevant passages. 

In the context of patent prior-art search, the query represents the \textbf{Citing Document} (specifically, its claims), while the retrieved targets are the \textbf{Cited Documents} (prior art). As illustrated below, patent queries exhibit extreme variance in length and contain highly specialized terminology. The examples also explicitly display the number of citing units (claims) and the number of cited units (ground-truth relevant documents and passages), highlighting the precise, fine-grained, and highly imbalanced nature of this retrieval task.

\vspace{1.5em}
\hrule\vspace{0.8em}
\noindent\textbf{Topic: PSG-7} \hfill \textit{Source (Citing Doc): EP-1246167-A1}\\
\textbf{IPC:} G10L 21/02 \quad \textbf{Citing Units:} 1 Claim (Query length: 79 words)\\
\textbf{Cited Units:} 2 Relevant Documents containing 12 Relevant Passages
\vspace{0.8em}\hrule\vspace{0.8em}

\noindent\textbf{QUERY (Patent Claim):} \\
1. An arrangement for activating and deactivating automatic noise cancellation (ANC) in a mobile station, in which there is an ANC facility (3) for cancelling background noise of an incoming audio signal and circuits for activating and deactivating ANC, characterized in that the arrangement includes an automatic circuit configuration (10) for detecting a need for ANC, according to a selected criterion, which detection circuit configuration (10) is arranged to automatically control the activation and deactivation of the ANC facility (3).

\vspace{0.8em}\hrule\vspace{0.8em}
\noindent\textbf{GROUND-TRUTH RELEVANT PASSAGES (Showing 3 of 12):}

\vspace{0.5em}
\noindent\textbf{[1] Doc ID:} EP-0661903-A2 \hfill \textbf{XPath:} \texttt{/patent-document/description/p[16]} \\
\textit{Text (43 words):} where $C_i(n)$ ($-L \leq i \leq L$) are the tap coefficients of the first adaptive filter at a time $n$, and $(2L+1)$ is a number of taps. The replica of received voice signal $y_1(n)$ thus estimated is supplied to the first subtracter 53.

\vspace{0.5em}
\noindent\textbf{[3] Doc ID:} EP-0661903-A2 \hfill \textbf{XPath:} \texttt{/patent-document/description/p[17]} \\
\textit{Text (58 words):} The first adaptive filter 51 also updates the tap coefficients based on, for example, the Least Mean Square (LMS) algorithm shown in the following equation (2) when the control signal SW1 is in the high level, that is, when the received voice signal $S(n)$ exceeds the predetermined value. $c_i(n+1) = C_i(n) + 2\mu_1 \cdot e(n) \cdot S(n)$

\vspace{0.5em}
\noindent\textbf{[2] Doc ID:} EP-0661903-A2 \hfill \textbf{XPath:} \texttt{/patent-document/description/p[19]} \\
\textit{Text (23 words):} Further, the first adaptive filter 51 stops the update of the tap coefficients when the control signal SW1 is in a low level.

\vspace{0.8em}\hrule

\vspace{2.5em}
\hrule\vspace{0.8em}
\noindent\textbf{Topic: PSG-34} \hfill \textit{Source (Citing Doc): EP-1669402-A1}\\
\textbf{IPC:} C08L 23/08 \quad \textbf{Citing Units:} 7 Claims (Query length: 332 words)\\
\textbf{Cited Units:} 3 Relevant Documents containing 53 Relevant Passages
\vspace{0.8em}\hrule\vspace{0.8em}

\noindent\textbf{QUERY (Patent Claims):} \\
1. The use of a rubber composition for preparing cable connector seals, said rubber composition comprising:
an ethylene/$\alpha$-olefin/non-conjugated polyene copolymer [A];
an organopolysiloxane [B] having an average composition formula (I) of R1tSiO(4-t)/2 wherein R1 is an unsubstituted or substituted monovalent hydrocarbon group and t is a number ranging from 1.9 to 2.1; and
an SiH group-containing compound [C];
said copolymer [A] and said organopolysiloxane [B] having a weight ratio ([A]:[B]) of 100:0 to 5:95. \\
2. The use according to Claim 1,
wherein the ethylene/$\alpha$-olefin/non-conjugated polyene copolymer [A] has:
(i) a mass ratio of ethylene to an $\alpha$-olefin of 3 to 20 carbon atoms (ethylene/$\alpha$-olefin) of 35/65 to 95/5;
(ii) an iodine value of 0.5 to 50;
(iii) an intrinsic viscosity [$\eta$] of 0.01 to 5.0 dl/g as measured in decalin at 135$^\circ$C; and
(iv) constituent units of non-conjugated polyene derived from at least one norbornene compound represented by the following formula [I] or [II]:
wherein n is an integer of 0 to 10, R1 is a hydrogen atom or an alkyl group of 1 to 10 carbon atoms, and R2 is a hydrogen atom or an alkyl group of 1 to 5 carbon atoms;
wherein R3 is a hydrogen atom or an alkyl group of 1 to 10 carbon atoms. \\
3. The use according to Claim 1 or 2, wherein said rubber composition contains a catalyst [D] in addition to the ethylene/$\alpha$-olefin/non-conjugated polyene copolymer [A], the organopolysiloxane [B] and the SiH group-containing compound [C]. \\
4. The use according to Claim 3, wherein said rubber composition contains a reaction inhibitor [E] in addition to the ethylene/$\alpha$-olefin/non-conjugated polyene copolymer [A], the organopolysiloxane [B], the SiH group-containing compound [C] and the catalyst [D]. \\
5. A cable connector seal obtainable by crosslinking the rubber composition as defined in any one of Claims 1 to 4. \\
6. A cable connector seal according to Claim 5, wherein the cured product has a durometer A hardness of 45 or less. \\
7. An automotive cable connector including the cable connector seal of Claim 5 or 6.

\vspace{0.8em}\hrule\vspace{0.8em}
\noindent\textbf{GROUND-TRUTH RELEVANT PASSAGES (Showing 3 of 53):}

\vspace{0.5em}
\noindent\textbf{[1] Doc ID:} EP-0855426-A1 \hfill \textbf{XPath:} \texttt{/patent-document/claims/claim[1]} \\
\textit{Text (42 words):} A moisture-curable elastomer composition obtained by dynamic heat treatment of a mixture comprising (a) an ethylene-$\alpha$-olefin-unconjugated diene copolymer rubber,(b) a silicon-based crosslinking agent having two or more SiH groups within the molecule,(c) a hydrosilylation catalyst and(d) a hydrolyzable silane group-containing thermoplastic resin.

\vspace{0.5em}
\noindent\textbf{[2] Doc ID:} EP-1070746-A2 \hfill \textbf{XPath:} \texttt{/patent-document/description/p[18]} \\
\textit{Text (155 words):} A dichloromethane solution containing 10 weight\% solids was prepared by dissolving an acrylic resin with a softening point of 85$^\circ$C (Elvacite 2008 from DuPont) in dichloromethane. Using a dual-fluid nozzle, this dichloromethane solution and distilled water were continuously sprayed into a spray dryer, the dichloromethane solution at 100 cc/minute and the distilled water at 25 cc/minute using a hot nitrogen gas current as propellant. The temperature of the hot nitrogen current during this process was 80$^\circ$C and the pressure was 0.25 kg/cm$^2$ (0.025 MPa). The resulting hollow acrylic resin particulate was immersed for 24 hours in an aqueous solution of 100 parts distilled water and 1 part nonionic surfactant (ethylene oxide adduct of trimethylnonanol), and the hollow acrylic resin powder that floated was fractionated and collected. The resulting hollow acrylic resin powder had an average particle size of 20 $\mu$m, contained nitrogen in its interior space, and had an average shell thickness of 4 $\mu$m.

\vspace{0.5em}
\noindent\textbf{[3] Doc ID:} EP-1070746-A2 \hfill \textbf{XPath:} \texttt{/patent-document/description/p[24]} \\
\textit{Text (272 words):} The following were introduced into and mixed to homogeneity in a kneader mixer: 100 parts dimethylvinylsiloxy-endblocked dimethylsiloxane-methylvinylsiloxane copolymer gum (weight-average molecular weight = 500,000) composed of 99.85 mole\% dimethylsiloxane units and 0.15 mole\% methylvinylsiloxane units, 45 parts wet-process silica with a BET specific surface area of 130 m$^2$/g (Nipsil LP from Nippon Silica Kogyo Kabushiki Kaisha), and 3 parts hydroxyl-endblocked dimethylsiloxane oligomer with a viscosity of 50 mPa$\cdot$s. The resulting mixture was heated for 60 minutes at 175$^\circ$C to produce a silicone rubber base compound. A moldable fire-resistant silicone rubber sponge composition was then prepared by mixing the following to homogeneity on a two-roll mill into 100 parts of the silicone rubber base compound: 30 parts mica powder, 1 part trimethylsiloxy-endblocked dimethylsiloxane-methylhydrogensiloxane copolymer with a viscosity of 25 mPa$\cdot$s (molar ratio of silicon-bonded hydrogen in this component to vinyl in the gum = 3.3 : 1), sufficient chloroplatinic acid/divinyltetramethyldisiloxane complex to provide 4 ppm platinum metal, 0.03 part 1-ethynylcyclohexanol as cure inhibitor, and 1 part of the hollow silicone resin powder prepared as described in Reference Example 1. This moldable fire-resistant silicone rubber sponge composition was introduced into an extruder configured for molding gaskets and a fire-resistant gasket of this composition with a diameter of 3 cm was molded. This gasket was placed in a forced hot-air convection oven and heated for 3 minutes at 250$^\circ$C to produce a fire-resistant silicone rubber sponge gasket. This fire-resistant silicone rubber sponge gasket was cut with a knife and the diameter of the contained foam cells was measured. The cells diameters were in the range from 0.2 to 0.5 mm. The expansion ratio was 2.25-fold.

\vspace{0.8em}\hrule

\vspace{2.5em}
\hrule\vspace{0.8em}
\noindent\textbf{Topic: PSG-26} \hfill \textit{Source (Citing Doc): EP-1255091-A1}\\
\textbf{IPC:} G01C 21/20 \quad \textbf{Citing Units:} 8 Claims (Query length: 457 words)\\
\textbf{Cited Units:} 8 Relevant Documents containing 429 Relevant Passages
\vspace{0.8em}\hrule\vspace{0.8em}

\noindent\textbf{QUERY (Patent Claims):} \\
1. A curve approach control device for predicting and judging the approach of a curve based on information for a curve ahead for a vehicle and outputting prescribed control commands characterized by comprising: means for comparing and judging matching between curve information supplied by a navigation system and real curve information based on actual vehicle behavior; and means for stopping outputting of the control commands when it is judged that the curve information from the navigation system and the real curve information do not match. \\
2. The curve approach control device of claim 1, characterized in that curve information from the navigation system is represented by curve depth calculated from node data relating to a road, the real curve information is represented by a value arrived at by integrating a turning motion parameter of the vehicle, and when a difference between the curve depth and the integral value for the turning motion parameter is a threshold value or more, it is judged that the curve information from the navigation system and the real curve information do not match. \\
3. The curve approach control device of claim 2, characterized in that non-matching of polarities of the curve depth and the integral value of the turning motion parameter is taken as one condition for judging that the curve information from the navigation system and the real curve information do not match. \\
4. The curve approach control device of claim 1, characterized in that curve information from the navigation system is represented by curve curvature calculated from node data relating to a road, the real curve information is represented by curve curvature calculated from vehicle speed and a turning motion parameter for the vehicle, and when a difference between the curve curvatures is greater than or equal to a threshold value, it is judged that the curve information from the navigation system and the real curve information do not match. \\
5. The curve approach control device of claim 4, characterized in that non-matching of polarities of the curve curvatures is taken as one condition for judging that the curve information from the navigation system and the real curve information do not match. \\
6. The curve approach control device of claim 2 or claim 4, characterized in that the turning motion parameter is yaw rate or lateral acceleration. \\
7. The curve approach control device of any one of claims 1, 2 or 4, characterized in that halting of outputting of the control commands continues for a set distance, a set time, or until mismatching is resolved. \\
8. A vehicle equipped with the curve approach control device of any one of claims 1 to 7, characterized in that at least one of alarm control, deceleration control and steering control is executed based on a signal from the curve approach control device.

\vspace{0.8em}\hrule\vspace{0.8em}
\noindent\textbf{GROUND-TRUTH RELEVANT PASSAGES (Showing 3 of 429):}

\vspace{0.5em}
\noindent\textbf{[3] Doc ID:} EP-0819912-A2 \hfill \textbf{XPath:} \texttt{/patent-document/description/p[6]} \\
\textit{Text (63 words):} According to the present invention, the vehicle driving condition prediction device predicts a lateral acceleration of the vehicle which will occur when the vehicle is going into a curve ahead. This prediction device predicts an approaching vehicle speed at a curve approach point based on a predetermined acceleration/deceleration pattern, and predicts a lateral acceleration in a curve from a predicted approaching vehicle speed.

\vspace{0.5em}
\noindent\textbf{[2] Doc ID:} EP-0819912-A2 \hfill \textbf{XPath:} \texttt{/patent-document/description/p[8]} \\
\textit{Text (67 words):} The above-mentioned predetermined acceleration/deceleration pattern is formulated by taking into consideration the driving condition at this point in time. For example, by setting conditions so that the present acceleration or deceleration is maintained for a specified period of time, it becomes possible to make a prediction for the vehicle speed by taking the present driving condition into account. Thus, a suitable curve approach speed can be predicted.

\vspace{0.5em}
\noindent\textbf{[1] Doc ID:} EP-0819912-A2 \hfill \textbf{XPath:} \texttt{/patent-document/description/p[12]} \\
\textit{Text (95 words):} The present invention is characterized in that the speed limit of the road on which the vehicle is running is considered in the predetermined acceleration/deceleration pattern. The drivers are aware of the speed limits of ordinary roads, and probably drive at speeds close to the speed limits particularly when traveling through curvy roads. Therefore, by assuming that the driver drives at the speed limit and also arranging for a warning to be issued when the driver must not continue to drive as fast as this and must slow down, a warning can be given appropriately.

\vspace{0.8em}\hrule

\section{Analysis of Center Construction}
\label{app:center-analysis}
We analyze the properties of the k-center vocabularies constructed across three encoder models (PatentMap, PaECTER, and BERT-for-Patents), three span units (encoder tokens, noun phrases, and hybrid token--phrase units), and vocabulary sizes $V \in \{10\text{k}, 20\text{k}, \ldots, 50\text{k}\}$. After farthest-first traversal, we examine three characteristics: \emph{embedding compactness}, \emph{representation sparsity}, and \emph{posting-list balance}.

\begin{figure}[ht]
    \centering
    \includegraphics[width=\textwidth]{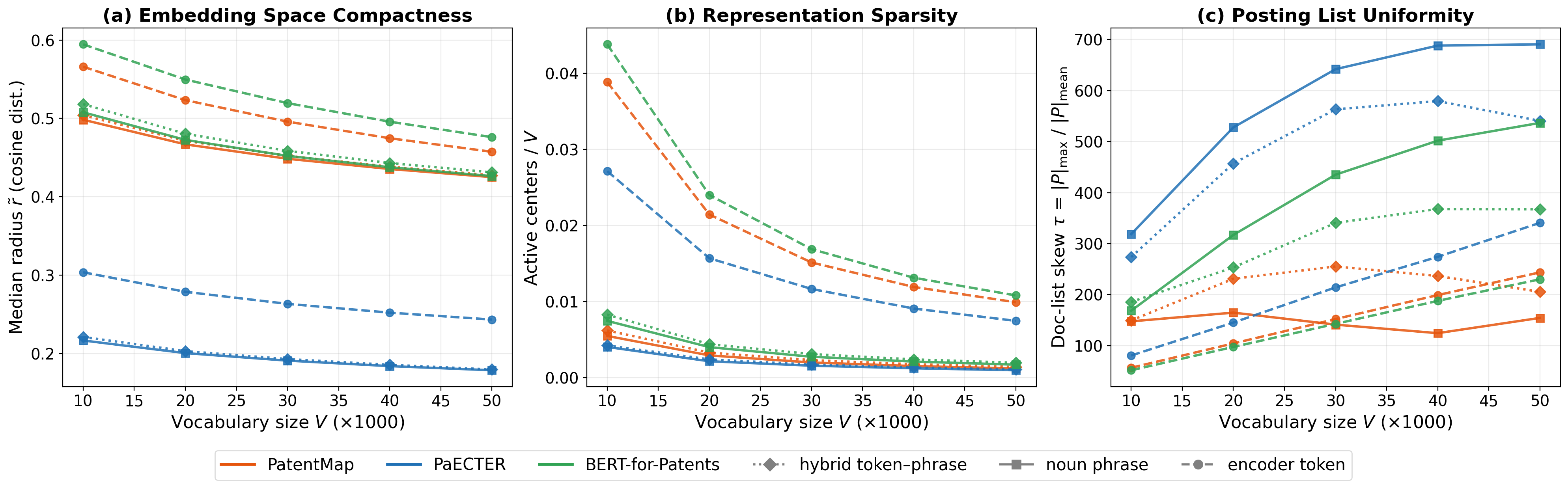}
    \caption{Properties of k-center vocabularies as a function of vocabulary size $V$.
    (a) Median Voronoi radius $\tilde{r}$; lower values indicate more compact local neighborhoods.
    (b) Mean number of activated centers per span, normalized by vocabulary size.
    (c) Posting-list skew $\tau = |P|_{\max}/|P|_{\mathrm{mean}}$; higher values indicate that a few centers dominate the index.
    Colors denote encoder models and line styles denote span units.} 
    \label{fig:kcenter-analysis}
\end{figure}

\paragraph{Embedding compactness.}
Figure~\ref{fig:kcenter-analysis}(a) shows that the median Voronoi radius $\tilde{r}$ decreases as the vocabulary size $V$ increases, confirming that larger vocabularies provide finer partitions of the span embedding space. The encoder choice also has a strong effect: PaECTER produces substantially smaller radii than PatentMap and BERT-for-Patents, indicating a more compact embedding geometry. Across encoders, encoder-token units are consistently the most dispersed, whereas noun-phrase and hybrid token--phrase units form more compact neighborhoods and often behave similarly. The decrease in $\tilde{r}$ is also steeper for encoder-token units than for noun-phrase and hybrid units. This suggests that token-level representations benefit more from increasing the vocabulary size: because individual contextual tokens occupy a broader and more heterogeneous embedding space, more centers are needed to cover them at fine resolution. 

\paragraph{Representation sparsity.}
Figure~\ref{fig:kcenter-analysis}(b) reports the mean number of activated centers per span, normalized by the vocabulary size.\footnote{The activation rate is measured before applying the Top-$K$ cap, using only the radius-based activation condition, in order to reflect the natural overlap of the learned center neighborhoods.} The activation rate decreases consistently as $V$ grows. This indicates that larger vocabularies not only refine the semantic partition but also make individual span representations sparser relative to the full vocabulary. Encoder-token units activate a larger fraction of centers than noun-phrase and hybrid units, especially for BERT-for-Patents and PatentMap, which is consistent with their more dispersed geometry in Figure~\ref{fig:kcenter-analysis}(a). For encoder-token units, PaECTER yields lower activation rates than the other encoders, which is consistent with its more compact geometry in Figure~\ref{fig:kcenter-analysis}(a). Overall, these results show that vocabulary size and span granularity jointly control the sparsity of Sparse Coverage representations.

\paragraph{Posting list balance.}
Figure~\ref{fig:kcenter-analysis}(c) examines posting-list skew using $\tau = |P|_{\max}/|P|_{\mathrm{mean}}$. A high value indicates that a small number of centers receive disproportionately long posting lists. The most important observation is that geometric compactness does not imply balanced posting lists. PaECTER, which has the most compact embedding space in Figure~\ref{fig:kcenter-analysis}(a), also exhibits the strongest skew overall, especially for noun-phrase units. BERT-for-Patents shows a similar but weaker pattern, with high skew for hybrid units. In contrast, PatentMap produces more balanced posting lists across span units and vocabulary sizes.

The skew also tends to increase with $V$. As the vocabulary becomes larger, the average posting-list length decreases, but generic or highly reusable semantic centers may still be activated by many documents. Their relative dominance therefore becomes more visible. This finding directly motivates the use of IDF weighting in Sparse Coverage: the k-center objective encourages coverage of the embedding space, but it does not control how often each center is activated across documents. Document-frequency calibration is therefore needed to reduce the influence of generic semantic regions during retrieval.

Overall, the analysis shows that the induced sparse vocabulary is shaped by both the encoder geometry and the span unit. Phrase-aware units, including noun phrases and hybrid token--phrase units, produce more compact local neighborhoods, whereas encoder-token units provide finer but more dispersed evidence. At the same time, compactness can increase posting-list imbalance, which explains why Sparse Coverage combines geometric center selection with Top-$K$ sparsification, stop-center suppression, and IDF-weighted retrieval.

\section{Additional Hyperparameter Ablations}
\label{sec:ablation_study}

We further analyze three hyperparameters of Sparse Coverage: the assignment strategy, the IDF exponent $\alpha$, and the corpus-side normalization exponent $\gamma$. The analysis is restricted to configurations for which complete sweeps are available: PaECTER and PatentMap-V0 with encoder-token and noun-phrase units, averaged over $V \in \{10k,20k,30k,40k,50k\}$. This gives 20 configurations per setting.  All sweeps vary one component at a time around the default configuration used in the main experiments: soft assignment with $K=5$, IDF exponent $\alpha=2.0$, and corpus-side span-count normalization with $\gamma=0.5$.

\begin{table}[t]
\centering
\caption{
Hyperparameter ablation for Sparse Coverage averaged over complete sweeps.
Each setting is averaged over PaECTER and PatentMap-V0, encoder-token and noun-phrase units, and $V \in \{10k,20k,30k,40k,50k\}$.
The default setting used in the main experiments is marked with $\star$.
}
\label{tab:hyperparam-summary}
\small
\setlength{\tabcolsep}{5pt}
\begin{tabular}{llccc}
\toprule
\textbf{Factor} & \textbf{Setting} & \textbf{Doc R@100} & \textbf{Doc mAP} & \textbf{MAP(D)} \\
\midrule
\multirow{5}{*}{Assignment}
& Hard & 85.09 & 32.21 & 23.21 \\
& Soft $K=3$ & 92.39 & \textbf{40.38} & 25.42 \\
& Soft $K=5$ $\star$ & \textbf{92.39} & 39.97 & 25.47 \\
& Soft $K=7$ & 91.68 & 39.68 & \textbf{25.51} \\
& Soft $K=10$ & 91.20 & 38.04 & 25.29 \\
\midrule
\multirow{5}{*}{IDF $\alpha$}
& 0 & 89.32 & 36.01 & 24.90 \\
& 1.0 & 91.40 & 38.49 & 25.44 \\
& 1.5 & 92.20 & 39.32 & 25.50 \\
& 2.0 $\star$ & 92.39 & 39.97 & 25.47 \\
& 3.0 & \textbf{93.79} & \textbf{40.43} & \textbf{25.70} \\
\midrule
\multirow{4}{*}{Norm. $\gamma$}
& 0 & 85.67 & 35.11 & 23.79 \\
& 0.5 $\star$ & \textbf{92.39} & \textbf{39.97} & \textbf{25.47} \\
& 1.0 & 65.01 & 14.93 & 17.45 \\
& 2.0 & 30.95 & 8.70 & 7.29 \\
\bottomrule
\end{tabular}
\end{table}

Table~\ref{tab:hyperparam-summary} shows that controlled overlap between center regions is important. Hard assignment, where each span is matched only to its nearest center, performs substantially worse, indicating that a purely Voronoi-style representation introduces excessive quantization error. Soft assignment in Sparse Coverage first activates centers whose radii cover the span and then applies a top-$K$ cap to prevent excessive overlap in dense regions. The gains saturate quickly: $K=3$ and $K=5$ perform similarly, while larger values do not consistently improve performance. This suggests that a small amount of radius-based overlap helps recover semantic matches across neighboring centers, but too much overlap can introduce noisy shared activations and longer posting lists.

IDF weighting also has a clear effect. Removing IDF weighting ($\alpha=0$) yields the weakest results in the sweep, which confirms that geometric coverage alone does not make all centers equally discriminative.  Increasing $\alpha$ improves the averaged results, with $\alpha=3.0$ performing best among the tested values.  In the main experiments, we keep $\alpha=2.0$ to use a single moderate setting across all encoders and span units rather than tuning this parameter for the evaluation set.

Finally, corpus-side span-count normalization has a strong effect, especially for encoder-token units. The default square-root normalization $\gamma=0.5$ gives the best average results. Without normalization ($\gamma=0$), retrieval quality drops on average, indicating that long corpus units can receive inflated scores simply because they contain more spans and therefore have more opportunities to activate query centers. This effect is mainly driven by encoder-token units, which are more sensitive to span-count bias; noun-phrase units are coarser and can be less affected by the absence of normalization. Stronger normalization is also harmful: $\gamma=1.0$ and $\gamma=2.0$ severely degrade performance, suggesting that over-penalizing long passages suppresses genuine semantic evidence. Overall, the ablations support the default configuration used in the main experiments: limited soft assignment, IDF-weighted scoring, and mild corpus-side length normalization.

\section{Full Sparse Coverage Results}
\label{app:full-results}

Table~\ref{tab:full-sparse-results} reports the full non-averaged Sparse Coverage results across all encoder, span-unit, and vocabulary-size combinations. These results underlie the averaged trends shown in Figure~\ref{fig:vocab-size}. They show that the best configuration depends on both the encoder and the span unit: encoder-token units tend to provide stronger ranking-oriented performance, while noun-phrase and hybrid units can achieve very high document-level recall in specific configurations.

\begin{table*}[ht]
\centering
\caption{
Full Sparse Coverage results on CLEF-IP 2013 across all encoders, span units, and vocabulary sizes.
Best results overall are highlighted in \textbf{bold}, and second-best overall are \underline{underlined}.
}
\label{tab:full-sparse-results}
\footnotesize
\setlength{\tabcolsep}{5pt}
\begin{tabular}{llcccccc}
\toprule
\multirow{2}{*}{\textbf{Encoder}} & \multirow{2}{*}{\textbf{Unit}} & \multirow{2}{*}{$V$} & \multicolumn{3}{c}{\textbf{Document-level}} & \multicolumn{2}{c}{\textbf{Passage-level}} \\
\cmidrule(lr){4-6} \cmidrule(lr){7-8}
& & & \textbf{Recall@100} & \textbf{mAP} & \textbf{PRES@100} & \textbf{mAP(D)} & \textbf{Recall@1000} \\
\midrule

\multirow{15}{*}{BERT-for-Patents}
& \multirow{5}{*}{enc.\ tok.}
& 10k & 85.90 & 40.29 & 72.22 & 26.80 & 39.21 \\
& & 20k & 95.21 & 46.26 & 84.33 & 27.57 & 44.08 \\
& & 30k & 95.21 & 46.70 & 83.46 & \underline{29.54} & 44.17 \\
& & 40k & 93.47 & 46.05 & 82.84 & \textbf{30.03} & 45.32 \\
& & 50k & 93.12 & 46.68 & 82.51 & 28.79 & 44.32 \\
\cmidrule(lr){2-8}
& \multirow{5}{*}{NP}
& 10k & 84.27 & 19.11 & 65.34 & 22.35 & 32.18 \\
& & 20k & 91.91 & 19.68 & 72.66 & 24.45 & 38.15 \\
& & 30k & 87.50 & 24.02 & 68.63 & 24.92 & 34.38 \\
& & 40k & 86.39 & 26.59 & 68.48 & 24.92 & 38.70 \\
& & 50k & 88.37 & 23.70 & 73.35 & 24.47 & 37.46 \\
\cmidrule(lr){2-8}
& \multirow{5}{*}{NP+tok.}
& 10k & 90.35 & 25.11 & 71.48 & 26.12 & 37.13 \\
& & 20k & 93.85 & 28.42 & 77.36 & 24.46 & 37.53 \\
& & 30k & 90.87 & 29.02 & 72.71 & 25.51 & 40.74 \\
& & 40k & 92.88 & 24.72 & 72.02 & 25.32 & 38.73 \\
& & 50k & 86.91 & 25.27 & 70.13 & 22.31 & 35.29 \\

\midrule

\multirow{15}{*}{PaECTER}
& \multirow{5}{*}{enc.\ tok.}
& 10k & 89.13 & 37.30 & 77.06 & 23.60 & 43.17 \\
& & 20k & 92.47 & 44.86 & 79.80 & 26.22 & 45.68 \\
& & 30k & 89.51 & 49.13 & 80.66 & 24.31 & 46.12 \\
& & 40k & 92.47 & \textbf{52.60} & 81.46 & 26.21 & \underline{48.60} \\
& & 50k & 93.58 & \underline{51.33} & 82.85 & 25.53 & \textbf{48.63} \\
\cmidrule(lr){2-8}
& \multirow{5}{*}{NP}
& 10k & 89.20 & 31.45 & 75.12 & 22.57 & 40.82 \\
& & 20k & 96.74 & 37.45 & 82.00 & 24.46 & 44.30 \\
& & 30k & 95.07 & 32.36 & 80.24 & 22.77 & 44.68 \\
& & 40k & 93.37 & 34.62 & 79.19 & 21.90 & 42.92 \\
& & 50k & 93.96 & 32.46 & 78.10 & 22.19 & 43.52 \\
\cmidrule(lr){2-8}
& \multirow{5}{*}{NP+tok.}
& 10k & 91.81 & 35.39 & 77.08 & 22.37 & 41.85 \\
& & 20k & 92.74 & 33.61 & 74.79 & 23.69 & 44.29 \\
& & 30k & 92.19 & 35.07 & 79.38 & 25.53 & 45.27 \\
& & 40k & 94.27 & 44.17 & \underline{84.38} & 25.31 & 45.68 \\
& & 50k & \textbf{99.31} & 45.44 & \textbf{85.21} & 25.35 & 45.37 \\

\midrule

\multirow{15}{*}{PatentMap-V0}
& \multirow{5}{*}{enc.\ tok.}
& 10k & 92.15 & 47.43 & 79.72 & 29.41 & 46.03 \\
& & 20k & 93.78 & 45.50 & 83.15 & 28.55 & 45.57 \\
& & 30k & 91.67 & 46.60 & 81.24 & 27.58 & 46.40 \\
& & 40k & 92.57 & 50.15 & 82.58 & 29.06 & 47.46 \\
& & 50k & 92.08 & 50.33 & 81.99 & 28.74 & 47.83 \\
\cmidrule(lr){2-8}
& \multirow{5}{*}{NP}
& 10k & 91.91 & 26.10 & 75.75 & 25.78 & 41.83 \\
& & 20k & \underline{97.71} & 30.90 & 81.29 & 25.40 & 43.52 \\
& & 30k & 88.72 & 31.54 & 76.84 & 24.46 & 42.51 \\
& & 40k & 88.58 & 35.28 & 74.93 & 25.09 & 42.55 \\
& & 50k & 93.06 & 31.98 & 80.37 & 25.59 & 42.33 \\
\cmidrule(lr){2-8}
& \multirow{5}{*}{NP+tok.}
& 10k & 95.76 & 28.00 & 79.00 & 27.77 & 40.74 \\
& & 20k & 92.26 & 33.15 & 77.15 & 25.52 & 42.89 \\
& & 30k & 95.03 & 34.99 & 78.41 & 25.35 & 44.02 \\
& & 40k & 95.00 & 34.38 & 81.18 & 24.86 & 44.64 \\
& & 50k & 97.26 & 34.22 & 80.31 & 26.23 & 45.40 \\

\bottomrule
\end{tabular}
\end{table*}


\end{document}